\documentstyle[11pt,aaspp4,epsfig]{article}

\def\simg{\mathrel{\hbox{\rlap{\lower.55ex \hbox {$\sim$}}
                   \kern-.3em \raise.4ex \hbox{$>$}}}}
\def\siml{\mathrel{\hbox{\rlap{\lower.55ex \hbox {$\sim$}}
                   \kern-.3em \raise.4ex \hbox{$<$}}}}

\def\Mesz{M\'esz\'aros~}
\def\beq{\begin{equation}}
\def\enq{\end{equation}}
\def\bea{\begin{eqnarray}}
\def\ena{\end{eqnarray}}

\def\bec{\begin{center}}
\def\enc{\end{center}}
\def\etal{{\it et al.}}

\def\cm2si{\hbox{cm$^{-2}$s$^{-1}$}}
\def\cmcui{{\rm cm}^{-3}}
\def\r13{r_{13}}
\def\L50{L_{50}}
\def\G25{\Gamma_{j2.5}}
\def\T7{T_7}
\def\Z26{Z_{26}}
\def\Gb{\Gamma_{b}}
\def\msun{M_\odot}
\def\eps{\epsilon}

\def\Lal{Ly$\alpha$~}
\def\Kal{K$\alpha$~}

\begin{document}

{\it \hfill ApJL, in press 5/31/03}

\title{Gamma-ray bursts as X-ray depth-gauges of the Universe}

\author{P. \Mesz$^{1}$ and M.J. Rees$^2$ }
\smallskip\noindent
$^1${Dpt.Astron. \& Astrophysics, Dpt. Physics, Pennsylvania State 
Univ., University Park, PA 16803}\\
$^2${Institute of Astronomy, University of Cambridge, Madingley Road, Cambridge
CB3 0HA, U.K.}


\begin{abstract}

We discuss the X-ray flux of gamma-ray burst afterglows at redshifts in 
the range 3-30, including the effects of the intergalactic He II absorption.
We point out that strong X-ray lines may form locally in burst afterglows 
starting minutes after the trigger. This can provide distinctive X-ray 
distance indicators out to the redshifts where the first generation of 
massive stars form.

\end{abstract}

\section{Introduction}

The first generation of stars in hierarchical cosmological models 
are expected to be metal-free and very massive (e.g.  Abel \etal 2002, 
Bromm \etal 2002). They evolve through a core collapse to form black 
holes of $M\sim 30-100\msun$ (Woosley, Zhang \& Heger 2003). These can 
produce, through accretion, a hard ionizing spectrum 
extending at least to soft X-rays. They may also lead to a first generation 
of 'collapsar' gamma-ray bursts (GRB), which would be about one order of 
magnitude more massive and luminous than their $z\sim 1-4$ counterparts 
(Schneider, Guetta \& Ferrara, 2002), with spectra extending beyond MeV 
energies. 
Spergel, \etal (2003) use WMAP polarization data to infer the post-reionization
optical depth to electron scattering, and deduce a reionization redshift
of $z \sim 17 \pm 5$. This deduction assumes  that at lower redshifts the
ionization is complete: if ionization happened gradually, then the heat
input would have to start at an even higher redshift to provide the
equvalent optical depth. 

Based on observations of the most distant quasars (e.g. Fan \etal, 
2001), below $z \sim 6.4$ the continuous Gunn-Peterson trough is replaced
by partial blanketing of the continuum by the Lyman $\alpha$ forest.
He II may play a significant role in determining the IGM opacity in the 
soft-X-ray range at $z \lesssim 6$ (Miralda-Escud\'e, 2000).  
The IR/optical/UV spectra of the first stars (Loeb \& Barkana 2002) and/or 
distant long GRB afterglows (Lamb \& Reichart, 2000, Ciardi \& Loeb 2001) 
can provide redshift markers through their HI \Lal cutoff.  For short GRB, 
however, optical afterglows may be weak or absent (Panaitescu \etal 2001).  
Hence, X-ray band redshift markers would be most useful, especially 
at $z\simg 6$, where pop. III objects and their remnants may dominate 
the ionizing output.

Here we discuss the role of the He II absorption, which, over most of
the volume of the high redshift IGM, is the major determinant of the 
optical depth bluewards of $54.4/(1+z)$ eV. The He II opacity becomes 
negligible at energies above $\sim 0.2$ keV, allowing the X-ray spectrum 
of high redshift sources to become observable in principle. The Fe group 
\Kal energy is in this range for objects at $z \lesssim 30$. These metals, 
while absent through most of the IGM at such redshifts, should be abundant 
in patches immediately surrounding pop. III objects, as well as in GRB.  
Other lines may also be useful at lower redshifts.  Here we discuss the 
role of the Fe-group \Kal X-ray lines as redshift markers in the range 
$z\sim 3-30$, including the first generation of GRB.

\section{IGM Opacities and GRB X-ray Fluxes}
\label{sec:fluxes}

At $z \simg 6.4$ the IGM appears to acquire an increasing fraction of
neutral gas (Fan, \etal, 2001), with the first protogalaxies and quasars 
presumably forming at $z \sim 6-9$. 
The cosmological paramaters derived from WMAP results are $\Omega_m= 0.29, 
\Omega_\Lambda=0.71, \Omega_b=0.047, h= 0.68$, with a lower limit on the 
reionization redshift $z_i \sim 17\pm 5$ (Spergel, \etal, 2003) which may 
signal the first generation of stars.
Bluewards of the HI Lyman limit the opacity 
drops with photon energy as $E^{-3}$. Helium is likely to be in its 
hydrogenic form, He II. At the energy 54.4 eV of the He II ionization 
edge the cross section is a factor $Z^4=16$ times higher than 
that of the H I bound-free cross section, and for a $\sim$ 10\% He 
abundance relative to H, the He II opacity will be at least a factor 1.6 
more important than that of H I, and continues to dominate at energies 
above that. Its influence should be even stronger, since He recombines at 
a rate 5.5 faster than H, provided the source spectra are not too hard 
and fewer photons are emitted at the He II ionization energy. 
The dominant sources at $z\sim 3$  are likely to satisfy this, so the 
far UV opacity is dominated by He II.  Miralda-Escud\'e (2001) estimates
this optical depth as $\tau_{0,HeII}\simeq 1.7\times 10^3 
(\Omega_b h Y/0.007) [H_o(1+z)^{3/2}/H(z)] [(1+z)/4]^{3/2}$, where $H(z)$ is
the Hubble expansion. Using the above parameters we obtain for the HeII edge 
optical depth at $z_c\sim 3$ 
\beq
\tau_{0,HeII} \simeq 4.3\times 10^3 \left({\Omega_b h Y \over 0.007}\right)
                     \left( {1+z_c} \over 4 \right)^{3/2}
\label{eq:tauHeII}
\enq
Above the edge the cross section is $\propto E^{-3}$ and the opacity 
becomes less than unity above 
\beq
E_t \sim 0.22 \left({\Omega_b h Y \over 0.007}\right)^{1/3} 
              \left( {1+z_c} \over 4 \right)^{-1/2}~\hbox{keV}.
\label{eq:Et}
\enq
which is the ``thining" photon energy observed at $z=0$ for which the IGM 
becomes transparent at the redshift $z_c$. Absorption in our galaxy 
can be comparable to that of intergalactic He II (Miralda-Escud\', 2000),
the energy (\ref{eq:Et}) corresponding to a galactic column density 
$\sim 2\times 10^{20}$ cm$^{-2}$ typical of moderately high latitudes.
Above this energy the intergalactic H opacity is also negligible, and 
X-ray sources at redshifts $z \simg z_c\sim 3$ become observable. 

The initial GRB luminosity is $L_{x,0}= 10^{50}L_{50}$ erg/s, or
$L_{E'}\sim 10^{49} (t'/t'_i)^{b} {E'}_{keV}^{a}$ erg/s/keV (0.2-10 keV),
where $t'_i\sim 10$ s is the time after trigger when the power law decay 
starts, $a\simeq 0.7$ is energy spectral index, $b_1\sim 1.1$ and 
$b_2\simeq 2$ are energy flux time decay indices before and after 
the lightcurve steepens at $t'_{br}\simeq 0.5$ day, primed quantities
being in the source frame and unprimed in the observer frame. The observer
frame energy flux of a source at $z$ at photon energy $E$ and time $t$ for 
a simple power law decay is
$
F_E(z,t)= { [L_{E,0} / 4\pi D_l(z)^2 (1+z)^{-1-a}]} 
         {[(t/t_i)^{b_1} / (1+z)^{b_1} ]} ~
          {\rm erg~cm}^{-2}{\rm s}^{-2}{\rm keV}^{-1}~,
$
while for a broken double power law in time it is
\beq
F_E(z,t)= {L_{E,0} (t_{br}/t_i)^{b_1}(t/t_{br})^{b_2} 
           \over 4\pi D_l(z)^2 (1+z)^{-1-a}}
 \left[{(t/t_{br})^{b_1-b_2}\over (1+z)^{b1}} H\left({{1+z}\over t/t_{br}}\right)
       + {1\over (1+z)^{b_2}} H\left({t/t_{br}\over {1+z}}\right) \right]~
         {\rm erg~cm}^{-2}{\rm s}^{-2}{\rm keV}^{-1}~,
\label{eq:Fxdouble}
\enq
where $H(x)$ is the step function,  $t_i= \min[t,10(1+z)]$ s 
and $t_{br}=5\times 10^4 (1+z)$ s are nominal values of the observer-frame 
decay initiation time and the break time, and $D_l$ is luminosity distance,
using $\Omega_{tot}=1$, $\Omega_m=0.3$, $\Omega_\Lambda=0.7$.
The observed X-ray flux of a nominal GRB afterglow at different observer 
times $t$  and redshifts $z$ derived from (\ref{eq:Fxdouble})
are shown in Table \ref{tab:1}.

\section {GRB X-ray Lines}
\label{sec:lines}

The strongest lines expected are from Fe \Kal at $E\sim 6.7/(1+z)$ keV, 
or possibly the nearby Ni, Co lines. Their detection requires high 
fluxes, which at high redshifts favors the earlier source times.
Of the five Fe lines in GRB reported from Beppo-SAX or Chandra, four were 
in emission at observer times $t \sim 0.5-1$ day, and one was in absorption 
at $t \sim 10-20$ s (e.g. Piro 2002). Lower Z metal lines were reported at 
$t\sim 8-12$ hours with XMM (Reeves \etal, 2002a; Watson \etal 2002) and 
Chandra (Butler, \etal, 2003). The most common interpretation of
the Fe lines is based on photoionized models. The absence so far of line 
detections at times between tens of seconds and $\sim$ 8 hours may in 
principle be due to current instrumental limitations on the time needed for 
slewing a narrow field X-ray detector. The Swift mission (Gehrels 2003), due 
for launch in December 2003, will have fast slewing capability ($\siml 1$ 
minute after trigger), and besides a wide field gamma-ray detector it is 
equipped with narrow field X-ray and optical detectors, both capable of 
moderate resolution spectroscopy. Thus, if strong enough lines are produced 
at minutes to hours, up to $\sim$ 100/year might be detected.
Line production is facilitated by a high gas density in the neighborhood of 
the GRB. In photoionized models, Fe line formation requires ionization 
parameters $\xi=L/n r^2 \sim 10^3$. This is easiest to satisfy, even at 
early times $\simg$ minutes, in ``local" models involving radii comparable 
to those of the progenitor stellar envelope.  

\noindent
{\it Stellar funnel models:} In these local models the lines involve
an envelope at 
$r\sim 10^{12}-10^{13}$ cm typical of giants (also for WR or He stars, this
radius is reached $10^3-10^4$ s after envelope expansion starts). The density 
is much higher than in supernova shells. The ionization could be due to a 
long-lasting jet (Rees \& \Mesz, 2000) or bubble (\Mesz \& Rees 2001), producing 
an X-ray continuum from shock accelerated electrons. X-ray absorption lines 
can arise while there is gas in front of the jet, and after it is swept away
the continuum from the jet on the sides of the funnel produces an emission 
reflection spectrum. The funnel wall gas density in pressure equilibrium with 
the jet is $n_{e,\ast}\sim 10^{21} \alpha \L50 \r13^{-2}\T7^{-1}\cmcui$, and 
the ionization parameter is $\xi=10^3 \alpha^{-1}\beta \T7$, where $\alpha$ 
is a geometrical factor and $\beta$ is ratio of ionizing to total luminosity. 
The recombination time of a hydrogenic ion of charge $Z=26\Z26$ is $t_{rec,Z}
\sim 2\times 10^{-10} \alpha^{-1}\L50^{-1}\r13^2 \T7^{3/2}\Z26^{-2}$ s. 
Balancing the number of ionizations $\beta L /4\pi r^2 \eps_i$, where 
$\eps_i\sim 10$ keV, with the number of recombinations 
$\delta n_e x_Z d_i/t_{rec,Z}$, where $x_Z$ is ion abundance ratio relative 
to H and $\delta$ allows for recombination to other ions, the depth from which
recombination photons escape is $ d_i \sim (\beta/\delta\alpha^2) \L50^{-1} 
\T7^{3/2}\r13^2 \Z26^{-2}x_Z^{-1}$ cm, and its Thomson scattering depth is
\beq
\tau_{i}\sim n_e \sigma_T d_i \sim 6\times 10^{-4} 
   (\beta/\delta\alpha) \T7^{1/2} \Z26^{-2} X_Z^{-1}~,
\label{eq:tauti}
\enq
which is $<1$ for an ion abundance ratio $x_Z > 6\times 10^{-4} 
(\beta/\alpha\delta) \T7^{1/2}\Z26^{-2}$. The solar Fe abundance ratio is 
$4\times 10^{-5}$, so an enrichment of $\simg 10$ above solar is needed for 
line photons to escape unscattered from $d_i$, which can arise from metals 
dragged in the jet and deposited in the walls.
Detailed radiative transfer calculations (Kallman, \Mesz \& Rees, 2003) for 
the shallow incidence angles of stellar funnel geometries and Fe abundances 
30-100 times above solar yield line equivalent widths $\sim 0.5-1.0$ keV. 
A difference in beaming of the line and continuum can be incorporated in the  
geometrical factor $\alpha$, wich affects the equivalent width. However, 
line photon reflections in the funnel (McLaughlin, Wijers, Brown \& Bethe 
2002) would not only lead Ni or Co photons to mimick Fe line photons, but 
would also tend to equalize the beaming factors of line and continuum. 
Note that the ionization parameter $\xi$ and the condition 
(\ref{eq:tauti}) for line photon emergence are independent of the 
luminosity (hence of the time) and the stellar radius. 

\noindent
{\it Jets with blobs:} X-ray lines can also be produced locally, inside the 
stellar funnel, by jets with blobs of entrained metal-rich material. 
The blobs will be initially subrelativistic, $\Gamma_{b} \sim 1$, with an 
equilibrium particle density $n_b\sim 10^{21}\L50\r13^{-2}\T7^{-1}\Gb^{-2}
\cmcui$, and an ionization parameter $\xi \sim L /n_b r^2 = 10^3 \T7\Gb^2$.
If the smoothed-out blob particle density is a fraction $\zeta$ of the 
jet density in the blob frame, ${\bar n}_b=\zeta n'_j(\Gamma_j/\Gamma_b)$,
where $n'_j\sim 10^{10}\L50\r13^{-2}\G25^{-2}\cmcui$ and $\Gamma_j=300\G25$
is the jet Lorentz factor, the blob volume filling 
factor is $f_v={\bar n}_b/n_b=3\times 10^{-9} \zeta \G25^{-1}\Gb \T7$, the
radius or smallest dimension of a blob is $r_b \sim (r/\Gamma_b)f_v/f_s$, 
where $f_s$ is the blob surface coverage factor (\Mesz \& Rees 1998). 
The Thomson depth of a blob is $\tau_{Tb}\sim 20\zeta\L50\r13^{-1} f_s^{-1}
\Gb^{-2}$, and at most a fraction $1/\tau_{Tb}$ of the ions undergo 
ionizations and recombinations contributing to the blob acceleration. The 
recombination time is  $t_{rec}\sim  4\times 10^{-10} \L50^{-1}\r13^2\T7^{3/2}
\Z26^{-2}\Gb^{2}$ s. A limit on the ratio between the time for accelerating a 
blob to $v\sim 0.1c$ cm/s and the stellar crossing time  is 
\beq
t_{ac}/t_{cr} \simeq t_{rec}(c/R_\ast) 56 A_{56} m_p 10^{-1}c \tau_{Tb} \sim 
5\times 10^{-3} \zeta f_s^{-1} \T7^{3/2} \Z26^{-2}A_{56}\G25^{-1}.
\label{eq:tacrec}
\enq
Another limit would apply after the blobs have accelerated and the density 
drops to the point where Thomson scattering becomes the main contributor to 
the acceleration, leading to 
\beq
t_{ac}/t_{cr} \sim 10^{-1}(R_\ast/r_s)(L/L_{Ed})^{-1}\tau_{Tb} \sim
 10^{-4}\zeta f_s^{-1} \G25^{-1},
\label{eq:tacsc}
\enq
where $r_s$ is Schwarzschild radius and $L_{Ed}$ is Eddington luminosity.
After becoming relativistic the blobs no longer contribute to  the
reprocessing, due to their lower density and higher ionization parameter, and 
their smoothed-out Thomson depth is at most comparable to that of the jet. 
E.g., a mist of blobs with $\zeta\sim 1$, $f_s\sim 1/3$ and $T\sim 
3\times 10^7$ has $t_{ac}/t_{cr}\sim 0.07$, and reprocesses the photoionizing 
continuum into narrow lines ($v/c\siml 0.1$) with a line to continuum ratio
$\siml 0.1$.

\noindent
{\it Transparency of local X-ray line models:} To observe 
unbroadened lines from the funnel walls or entrained blobs requires the jet to
be optically thin to both electron scattering and pair production. A jet of 
isotropic equivalent luminosity $L_{j,0} \sim 10^{50}L_{j,50}$ erg/s and 
Lorentz factor $\Gamma_j=300 \G25$ has a lab-frame proton density $n_p \sim 
3\times 10^{12} L_{50} r_{13}^{-2} \G25^{-2}$ cm$^{-3}$ at $r=10^{13}r_{13}$ 
cm. The scattering optical depths transverse and parallel to the jet axis are 
\beq
\tau_{\perp},\tau_{\parallel} \sim  2~ \L50 \r13^{-1} \G25^{-1}\theta_{-1}~,~
                                 2\times 10^{-4} \L50 \r13^{-1} \G25^{-3}~,
\label{eq:tausc}
\enq
where $\theta=10^{-1}\theta_{-1}$ is the jet opening angle. 
The internal shocks responsible for the $\gamma$-rays occur at radii 
$r_{sh,i}\sim 2c t_{var} \Gamma_j^2 \simg  10^{13}$ cm for variability 
timescale $t_{var}\simg 3 r_{13} \G25^{-2}$ ms. 
Hence the jet electrons and thermal photons are cold in the comoving 
jet-frame. The initial temperature at $r_o=10^7$ cm is 
$T_o\sim 0.5 L_{50}^{1/4}r_{o,7}^{-1/2}$ MeV. For
$T=T_o (r/r_o\Gamma_j)^{-2/3}$ (\Mesz \& Rees, 2000), at $10^{13}$ cm 
the thermal jet photon  energy is $\eps_{\gamma,j}\sim 3 \L50^{1/4} 
r_{07}^{1/6} \G25^{2/3} r_{13}^{-2/3}$ keV. The corresponding photon 
density near the spectral peak is $n_{\gamma,j}\sim 3\times 10^{20} 
\L50^{3/4} r_{13}^{-4/3} r_{o,7}^{-1/6} \G25^{-2/3}$ cm$^{-3}$ in the lab. 
The jet photons interact with X-ray photons reflected or generated by 
shocks at the funnel walls of energy $\eps_{\ast,ref}\simg 1 \L50^{1/4} 
r_{13}^{-2}$ keV, and with wall thermal photons of $\siml 0.1$ keV.  
Jet photons which produce $e^\pm$ via $\gamma\gamma$ must have an energy 
$\eps_{j,thr}\simg (m_e c^2)^2 /\eps_{\ast,ref} \sim 250$ MeV.
For a purely blackbody jet spectrum the number of photons at
this energy is exponentially small. For a comptonized tail above the 
spectral peak, assuming the flattest plausible slope $n_\gamma 
\propto \eps^{-2}$ we have at threshold a jet photon density 
$n_{j,thr}\sim 10^{10}L_{50}^{7/4} r_{13}^{-20/3} r_{o,7}^{1/6}\G25^{2/3}$ 
cm$^{-3}$, and the $\gamma\gamma$ optical depth is 
\beq
\tau_{\gamma\gamma}\siml 10^{-2}L_{50}^{7/4}r_{13}^{-17/3}\theta^{-1}
r_{o,7}^{1/6}\G25^{2/3}~.
\label{eq:taugg}
\enq
The opacities (\ref{eq:tausc}) and (\ref{eq:taugg}) are upper limits
for the initial jet values, which would drop in time.  Hence already at 
early times the jet is transparent to X-ray line photons. From the above 
scalings, such lines can escape unbroadened even at early times of minutes, 
when $L_x\sim 10^{50}$ erg/s, and extending to $t \simg$ day. 

\noindent
{\it Supranova model:} This non-local scenario for X-ray line formation
invokes a supernova which occurs $\sim$ months before  
the GRB (Vietri \& Stella, 1999, Lazzati, \etal 1999, Weth, \etal, 2000, Vietri 
\etal, 2001). It attributes the X-ray lines starting at $\sim$ 1 day 
to the geometrical light-travel time to a SN shell at $\simg 10^{16}$ cm. 
Recent data on GRB030329 (Henden, \etal 2003; Stanek, \etal 2003) 
indicate that a supernova occurred within $\sim 0-2$ days of the GRB, 
not enough for a shell to reach $\sim 10^{16}$ cm. 
The supranova geometry is not conducive to emission lines at early 
(minutes-hours) times, although absorption lines might in principle be 
expected early. However, for significant equivalent widths, especially in 
emission, the shell must be extremely clumpy to have sufficiently short 
recombination times. At $z \simg 6$ one would require earlier source times 
and higher incident luminosities than at low redshifts, and therefore even 
more extreme clumpiness factors, and smaller coverage factors. 
Radiative transfer calculations (Kallman, \etal, 2003) indicate that large 
equivalent widths are favored by shallow incidence angles, which are not 
generally expected in supernova shells, while being a natural feature of 
the ``local" stellar funnels above.  
In any case, the large continuum X-ray fluxes of equation (\ref{eq:Fxdouble}) 
and Table 1 are independent of the specific line mechanism, the requirement 
for detecting lines at high $z$ being a mechanism which produces high 
equivalent widths, preferably at early (minutes to hours) source times.  
There is one early detection report 10 s after a GRB trigger, in absorption 
(Amati, \etal, 2001), which is unusual, since so far sensitive X-ray 
measurements were started only several hours after the trigger.

\noindent
{\it High redshift X-ray fluxes and lines:}
Pop. III stellar masses are larger and core temperatures are higher 
than in lower redshift stars, but the external envelope radii may be 
comparable. The core collapse of pop. III objects would lead to a more 
massive black hole and could produce higher luminosity GRB (Schneider 
\etal, 2002). However funnel metal enrichment and X-ray line reprocessing 
would occur in the same way as for low redshift GRB, with similar 
equivalent widths. In Table \ref{tab:1} the pop. III GRB might appear
at $z \simg 6.5$, and it is likely that for these the scaling $L_{x,50}$ 
used for the initial continuum X-ray fluxes is an underestimate, which 
may need to be increased by a factor $\simg 10-30$.
The fourth column of table \ref{tab:1} gives the Fe \Kal line energies in the
observer frame, for each redshift. The Ni and Co line energies would be within 
$\sim 10\%$ of these. At $z=30$ the lines appear at the threshold where 
the He II optical thickness reaches unity, and for $z \siml 20$ they 
are well within the regime where the continuum is unabsorbed by the IGM.

In some bursts (Reeves \etal, 2002; c.f Rutledge \& Sako 2003;  Watson \etal, 
2002, Butler \etal, 2003) Si, S, Mg etc lines are reported, but not Fe. This may be due to a steeper photoionizing continuum (Lazzati, Ramirez-Ruiz \& Rees, 
2002). Such lower Z metal lines have generally lower strength and several  
closely grouped lines are expected together. 

\section{Discussion}
\label{sec:disc}

Studies of the first structures and the intergalactic medium at 
$z \simg 6$ require very bright sources with good distance indicators.  
Above these redshifts quasars are extremely rare, 
and the most numerous luminous sources expected  are protogalaxies, 
intermediate mass accreting black holes and supernovae, as well as GRB.
The latter are an outcome of massive stellar evolution, which is likely 
to precede galaxy formation and could have started at redshifts as high as 
$z\sim 10-30$ (e.g. Bromm \& Loeb, 2002).
Within the first hours in its rest frame ($\sim$ day in observer frame), 
the GRB afterglow brightness exceeds that of the most luminous AGNs or 
any other quasi-steady source.  This is particularly interesting in the 
X-ray range, where the He II Gunn-Peterson absorption is unimportant at 
energies $\simg$ 0.2 keV for sources at $z\simg$ 4. With space missions
such as Swift, pointed X-ray observations started within minutes can 
achieve detections out to $z \sim 10-30$ in integration times of hours, 
if GRB occur at those redshifts.

Here we have discussed the possibility of measuring GRB redshifts in the
range $z\simg 3-30$ using Fe group \Kal lines or edges. These lines
may form at early (minutes to days) times, independently of whether a 
supernova shell is produced. The observed line energies shown in 
Table \ref{tab:1} are well above the IGM He II opacity cut-off for 
redshifts $\simg 20$. Lines from lower Z metals can also be used, 
which fall below the He II cutoff at lower redshifts than Fe, e.g., 
S XVI \Kal at $\sim 2.6/(1+z)$ is useful only out to $z\siml 11$. 
X-ray redshift measurements are especially interesting, since for detecting 
H \Lal cut-offs in faint distant objects NIR spectroscopy is necessary, 
mostly relying on large area ground telescopes subject to observing 
conditions and scheduling. 

For GRB similar to those observed at $z\siml 4$ the nominal X-ray fluxes 
(Table 1) at $z \siml 12$ are bright enough for low resolution spectroscopy 
with Swift for $t \siml 10^3-10^4$ s, and easily detectable at least for 
$t \simg 10^5$ s. Because of the increasing K-corrections the fluxes do not
decrease much with redshift beyond $z\sim 4-5$ and saturate or turn up near 
their values for $z\sim 8-12$. For such nominal GRB, the fraction of the 
total number expected at $z\simg 5$ is $\simg 50\%$, but only $\sim 15\%$ may 
be detectable in flux-limited surveys, e.g. with Swift (Bromm \& Loeb 2002).  
However, at $z\simg 6-10$ the first generation of (pop. III) stars is likely
to be very massive, the resulting black holes from core collapse could 
have masses $10-30~M_\odot$, and if these lead to GRB, their luminosities
could be significantly higher than at low redshifts, with factors $L_{E,0}$ 
in equations (\ref{eq:Fxdouble}) which could be 10-30 times higher than 
assumed in Table 1. In this case, the fraction at $z\simg 5$ detectable 
in Swift's flux-limited survey could be $\simg 20-30\%$ of the total, 
or $\simg 20$/year.

With Chandra or XMM, grating spectroscopic observations of X-ray lines with 
rest-frame equivalent widths $\sim 0.5$ keV (as reported in nearby bursts) 
appear possible out to at least $z\simg 12$.  For instance, for the 
``nominal" GRB of Table 1 of initial $L_{x,0}=10^{50}$ erg/s, with a line 
at $E_{obs} \sim 0.5$ keV of equivalent width $EW_{obs}= EW_{rest}/(1+z)
\sim 0.04$ keV, observed at $t_{obs}=10^5$ s and $z=12$, a simulation with 
the XMM epic and XSPEC software (N. Brandt, private communication) with an 
integration time $10^5$ s gives a fit for a power law spectrum plus a 
Gaussian line of $\chi^2=682$ for 679 degrees of freedom, while a power 
law fit without the line gives $\chi^2=642$ for 676 degrees of freedom, 
corresponding to a line detection probability $\simg 99.9\%$. 
A K-edge might be expected at $\sim (4/3)E_{K\alpha}$, which could 
further improve the detection probability. For an initial $L_{x,0}=10^{51}$ 
erg/s, the fits should improve, and may extend to higher redshifts.  
If no O/IR redshifts are available, the presence of just one feature 
(or at most two, in the energy ratio 1:4/3) in the range 0.2-1 keV would 
strongly suggest that these are due to Fe-group elements, constraining the 
redshift to $\sim \pm 10\%$. Confusion with S, Si, Mg etc can be avoided, 
since these would appear as multiple lines, possibly with Fe lines at 
energies a factor $\sim 2.5$ higher. An isolated line or edge from Fe or 
Fe-group elements (\S \ref{sec:lines}) can be expected in GRB resulting from 
massive stellar collapses, where these spectral features arise in the dense 
outer envelope of the progenitor.  If pre-GRB supernova shells exist and if 
they generate similar equivalent widths, the same detection probabilities 
would apply.  With an expected GRB detection rate of 100-150/year by Swift 
and onboard X-ray follow-up capabilities starting minutes after the trigger, 
significant progress may be possible in the investigation of the 
first stars and the high redshift universe.

\acknowledgements
This research has been supported through NASA NAG5-9192, NAG5-9153
and the Royal Society.  We are grateful to L-J Gou, T. Abel, N. Brandt 
and D. Lazzati for useful discussions, and to the referee for valuable
comments.

\begin{table}
\caption{GRB afterglow properties in the observer frame for different redshifts.
The columns give the redshift; the intergalactic H Ly$\alpha$ wavelenght,
bluewards of which the Gunn-Peterson absorption sets in; the He II ``thinning" 
energy $E_t$, bluewards of which intergalactic absorption becomes negligible
and the continuum reemerges from the through; the Fe \Kal line energy;
and the GRB afterglow continuum X-ray flux from equation (\ref{eq:Fxdouble})
at six different observer times $t$ after the trigger (in erg/cm$^{2}$/s/keV). 
A decay $\propto t^{-1.1}$ is assumed after $t_i=10(1+z)$ s, steepening to 
$t^{-2}$ after $t_{br}=0.5(1+z)$ days, for an initial source-frame 0.2-10 keV 
luminosity normalized to $L_{x}=10^{50}$ erg/s at $t_i$. The latter is likely 
to be an underestimate at $z\simg 6.5$, for which the fluxes might be 10-30 
times higher than the values given in the table.}
\bigskip
\begin{tabular}{rrrrrrrrrr}
\hline
\hline
$z$ & ${\lambda_{\rm Ly\alpha,H} \over \mu{\rm m}}$ & ${E_t\over {\rm keV}}$ 
  & ${E_{Fe,{\rm K}\alpha}\over {\rm keV}}$
  & $F_E(10{\rm s})$ & $F_E(10^2{\rm s})$ & $F_E(10^3{\rm s})$ 
  & $F_E(10^4{\rm s})$ & $F_E(10^5{\rm s})$ & $F_E(10^6{\rm s})$
\\
\hline
\smallskip
3   & 0.486 & 0.22 & 1.675 &$ 1.9^{-9 } $&$ 6.8^{-10}$&$ 5.4^{-11}$&$ 4.3^{-12}$&$ 3.4^{-13}$&$ 2.2^{-14}$\\
6.5 & 0.912 & 0.22 & 0.893 &$ 6.1^{-10} $&$ 4.4^{-10}$&$ 3.5^{-11}$&$ 2.8^{-12}$&$ 2.2^{-13}$&$ 1.8^{-14}$\\
9.0 & 1.216 & 0.22 & 0.670 &$ 4.1^{-10} $&$ 4.1^{-10}$&$ 3.3^{-11}$&$ 2.6^{-12}$&$ 2.1^{-13}$&$ 1.6^{-14}$\\
12  & 1.581 & 0.22 & 0.515 &$ 3.0^{-10} $&$ 3.0^{-10}$&$ 3.2^{-11}$&$ 2.5^{-12}$&$ 2.0^{-13}$&$ 1.6^{-14}$\\
18  & 2.310 & 0.22 & 0.353 &$ 2.0^{-10} $&$ 2.0^{-10}$&$ 3.2^{-11}$&$ 2.6^{-12}$&$ 2.1^{-13}$&$ 1.6^{-14}$\\
30  & 3.770 & 0.22 & 0.216 &$ 1.3^{-10} $&$ 1.3^{-10}$&$ 3.5^{-11}$&$ 2.8^{-12}$&$ 2.2^{-13}$&$ 1.8^{-14}$\\
\hline
\label{tab:1}
\end{tabular}
\end{table}

\end{document}